\documentclass[conference]{IEEEtran}
\IEEEoverridecommandlockouts
\usepackage{cite}
\usepackage{amsmath,amssymb,amsfonts}
\usepackage{algorithmic}
\usepackage{graphicx}
\usepackage{textcomp}
\usepackage{xcolor}
\usepackage{longtable}
\usepackage{lscape}
\usepackage{array} 
\usepackage{float}
\usepackage{graphicx}
\usepackage{multirow} 
\usepackage{array} 
\def\BibTeX{{\rm B\kern-.05em{\sc i\kern-.025em b}\kern-.08em
    T\kern-.1667em\lower.7ex\hbox{E}\kern-.125emX}}
\begin{document}

\title{Leveraging Large Language Models for Cybersecurity: Enhancing SMS Spam Detection with Robust and Context-Aware Text Classification}

\author{\IEEEauthorblockN{1\textsuperscript{st} Mohsen Ahmadi}
\IEEEauthorblockA{\textit{Dept. Electrical Engineering and Computer Science} \\
\textit{Florida Atlantic University}\\
FL, USA \\
mahmadi2021@fau.edu}
\textit{Corresponding author}\\
\and
\IEEEauthorblockN{2\textsuperscript{nd} Matin Khajavi}
\IEEEauthorblockA{\textit{Foster School of Businesses} \\
\textit{ University of Washington}\\
Washington, USA\\
}
\and
\IEEEauthorblockN{3\textsuperscript{nd} Abbas Varmaghani}
\IEEEauthorblockA{\textit{Dept. Computer Engineering} \\
\textit{Islamic Azad University of Hamadan}\\
Hamedan, Iran\\
}
\and
\IEEEauthorblockN{4\textsuperscript{rd} Ali Ala}
\IEEEauthorblockA{\textit{Dept. Mechanical Engineering}\\
\textit{University College Dublin}\\
Dublin, Ireland\\
}
\and
\IEEEauthorblockN{5\textsuperscript{nd} Kasra Danesh}
\IEEEauthorblockA{\textit{Dept.Electrical and Computer Science} \\
\textit{Florida Atlantic University}\\
FL, USA \\
}
\and
\IEEEauthorblockN{6\textsuperscript{th} Danial Javaheri}
\IEEEauthorblockA{\textit{Dept. Electrical and Computer Science} \\
\textit{Korea University}\\
Seoul 02841, Republic of Korea \\
}
}
\maketitle

\maketitle

\begin{abstract}
This study evaluates the effectiveness of different feature extraction techniques and classification algorithms in detecting spam messages within SMS data. We analyzed six classifiers—Naive Bayes, K-Nearest Neighbors, Support Vector Machines, Linear Discriminant Analysis, Decision Trees, and Deep Neural Networks—using two feature extraction methods: bag-of-words and TF-IDF. The primary objective was to determine the most effective classifier-feature combination for SMS spam detection. Our research offers two main contributions: first, by systematically examining various classifier and feature extraction pairings, and second, by empirically evaluating their ability to distinguish spam messages. Our results demonstrate that the TF-IDF method consistently outperforms the bag-of-words approach across all six classifiers. Specifically, Naive Bayes with TF-IDF achieved the highest accuracy of 96.2\%, with a precision of 0.976 for non-spam and 0.754 for spam messages. Similarly, Support Vector Machines with TF-IDF exhibited an accuracy of 94.5\%, with a precision of 0.926 for non-spam and 0.891 for spam. Deep Neural Networks using TF-IDF yielded an accuracy of 91.0\%, with a recall of 0.991 for non-spam and 0.415 for spam messages. In contrast, classifiers such as K-Nearest Neighbors, Linear Discriminant Analysis, and Decision Trees showed weaker performance, regardless of the feature extraction method employed. Furthermore, we observed substantial variability in classifier effectiveness depending on the chosen feature extraction technique. Our findings emphasize the significance of feature selection in SMS spam detection and suggest that TF-IDF, when paired with Naive Bayes, Support Vector Machines, or Deep Neural Networks, provides the most reliable performance. These insights offer a foundation for improving SMS spam detection through optimized feature extraction and classification approaches, particularly in the context of large language models (LLMs), subversive security threats, and traditional machine learning methodologies.
\end{abstract}

\begin{IEEEkeywords}
Feature Extraction, Machine Learning, Spam Detection, Large Language Models (LLM), Subversive Security, Traditional Modeling, Support Vector Machines, Deep Neural Networks, Natural Language Processing.
\end{IEEEkeywords}

\section{Introduction}
Natural language processing (NLP) has traditionally concentrated on text classification, where various text elements like phrases, questions, paragraphs, and entire documents are categorized or labeled. Text classification serves a plethora of purposes, from spam detection to sentiment analysis, across different sectors, including email filtering, automated question answering, content categorization, and more [1-3]. Textual data is derived from numerous platforms such as websites, emails, messaging apps, social networks, support tickets, and customer feedback, among others. Despite being a rich resource, the inherent unstructured format of text presents challenges for efficient data mining and insight extraction. Text classification can be manual or automated, but with the growing bulk of the text in commercial use, the latter is gaining more importance. Automatic text classification can be broadly categorized into rule-based systems and machine-learning approaches. Rule-based systems operate on fixed criteria and require extensive domain knowledge, while machine learning models learn from examples to determine patterns for classification [4,5].

Supervised machine learning has been widely examined within text classification and information retrieval, aiding in the distinction and comprehension of various concepts. Traditional models employ a two-step process involving feature extraction (e.g., BoW, TfidfVectorizer) followed by classification using algorithms like Naive Bayes, SVMs, DNNs, Decision Trees, KNN, and LDA. While manual text categorization is expertise-dependent and scales poorly, the two-step machine-learning process can struggle with generalization and is often restricted by the need for extensive feature engineering [6]. However, while predefined features limit model performance, supervised machine learning and neural networks have proven capable of achieving high accuracy, provided they are trained with sufficient labeled data. On the other hand, unsupervised methods like hierarchical clustering and K-means, which do not rely on labeled data, are less accurate and more challenging to apply to large and imbalanced datasets [7,8]. Addressing the classification of underrepresented classes remains a significant hurdle due to dataset imbalances, which is a common real-world issue. There is ongoing research to develop sophisticated classifiers and enhance text representation. Text categorization streamlines the differentiation and systematic arrangement of text based on thematic content, which in turn improves search outcomes and user interactions with key datasets. Our research marks a significant progression from the traditional BoW approach to adopting the TF-IDF method, moving the field forward in terms of accuracy and efficiency.

The TF-IDF technique is an advanced feature extraction tool that computes values for each term within a document, elevating the significance of terms that are unique to a specific document while diminishing the importance of terms common across various documents. This approach enhances the representation of text data, making it both more informative and discriminating for analysis purposes. Additionally, we utilized principal component analysis (PCA) to streamline the feature set. PCA is a statistical procedure in data analysis and machine learning that concentrates on reducing dimensionality while retaining as much of the data's variability as possible. It is particularly beneficial for processing high-dimensional datasets, which can be cumbersome due to their vast number of features. By condensing the feature space, PCA not only eases computational demands but also potentially boosts the efficacy of certain machine learning models.

This study also delves into several prevalent supervised machine learning techniques in current research circles, including Naive Bayes (NB), Support Vector Machines (SVM), k-nearest Neighbors (KNN), Deep Neural Networks (DNN), Decision Trees (DT), and Latent Dirichlet Allocation (LDA). We evaluate the impact that each of these algorithms has on the task of text classification. Given the critical role that text classification plays in a wide array of applications, the quest to refine these algorithms remains a vibrant field of research.

\section{Related Work}
In recent times, there has been a surge of comprehensive reviews and analytical studies in the field of text classification within NLP. These reviews span a gamut of methodologies, from conventional linguistic approaches to the latest deep learning techniques. For instance, Roy et al. [10] have effectively applied LSTM and CNN models to the challenge of filtering spam in SMS messages, although their models were primarily optimized for texts in English. Another study [11] advocates for the adoption of various supervised machine-learning models for differentiating between spam and legitimate messages. In this study, a comparative analysis among Naive Bayes, maximum entropy, and SVM classifiers revealed that SVMs outperformed the others with an accuracy of 97.4\% on datasets in real-time environments, albeit at the cost of higher memory usage.

\begin{table*}
\centering
\caption{Summary of Recent Research on Spam Detection and Related Areas}
\label{tab:research_summary}
\begin{tabular}{|p{3cm}|p{1cm}|p{5cm}|p{3.5cm}|p{4cm}|}
\hline
\small
\textbf{Author} & \textbf{Year} & \textbf{Aim} & \textbf{Method} & \textbf{Result} \\
\hline
Shen et al. [45] & 2025 & Leverage BERT and GCN to improve SMS spam detection. & BERT-G3CN & Achieved high accuracy (99.28\% and 93.78\%) in benchmark datasets. \\ \hline
Haider Rizvi et al. [46] & 2025 & Provide a comprehensive review of GCN-based text classification methods. & GCN-based Text Classification & Summarized key strengths, challenges, and research gaps in GCN approaches. \\ \hline
Tusher et al. [47] & 2025 & Analyze deep learning methods for filtering email spam. & Deep Learning & Outlined various deep learning methods, their effectiveness, and future research needs. \\ \hline
Gong et al. [48] & 2025 & Survey existing research on language models for code optimization. & Systematic Literature Review & Identified research gaps and challenges in LM-based code optimization. \\ \hline
Liu et al. [49] & 2024 & Review applications of PLMs in cybersecurity. & PLM Analysis & Explored various cybersecurity applications of PLMs and future research directions. \\ \hline
Chen et al. [50] & 2024 & Investigate how LLMs enhance threat detection in cybersecurity. & LLM-based Threat Detection & Reviewed use cases, limitations, and potential improvements for LLM-based threat detection. \\ \hline
Alshatnawi et al. [51] & 2024 & Utilize contextual word embeddings to refine social media spam detection. & BERT, ELMo & Achieved superior accuracy using ELMo embeddings in detecting social media spam. \\ \hline
Chataut et al. [52] & 2024 & Compare the effectiveness of ML models and LLMs in spam detection. & Traditional ML vs. LLMs & LLMs demonstrated better performance in spam detection compared to traditional ML models. \\ \hline
Raja Abdul et al. [58] & 2024 & Identify smishing attacks using ML and NLP techniques. & SMSecure & Showed improved accuracy in identifying smishing attacks. \\ \hline
Raja Abdul Samad et al. [54] & 2024 & Propose an advanced model integrating phonetic and textual embeddings for detecting Chinese spam. & BiGRU-TextCNN & Enhanced Chinese spam detection by incorporating phonetic embeddings. \\ \hline
Shrestha [55] & 2023 & Introduce a refined approach to detecting smishing attacks with machine learning. & Random Forest, Extreme Gradient Boosting & Machine learning models exhibited high accuracy in detecting smishing attempts. \\ \hline
Yao et al. [56] & 2022 & Develop a high-performance spam email detection model. & XLNet & XLNet-based model outperformed existing methods with high precision and recall. \\ \hline
\end{tabular}
\end{table*}

Further developments in the domain have been marked by [12, 13], who introduced new approaches for spam SMS filtering utilizing LSTMs and RNNs within the frameworks of Keras and TensorFlow. This resulted in an impressive accuracy rate of 98\% when tested on the UCI dataset, yet the complexity of these models was considerably high due to their intricate architectures.  Lee and Kang [14] developed an SMS spam filtering method that utilizes a feed-forward neural network in combination with CBOW word embeddings. Their study revealed that simply increasing the number of hidden layers—such as from 27 onward—did not significantly enhance the accuracy of the model. This emphasizes the importance of layer quality over sheer quantity. The research landscape has been further advanced by computer scientists concentrating on machine learning and deep learning-based models. For instance, Xu et al. [15] introduced a novel technique to extract textual features from email image attributes and employed SVM for classification to optimize spam filtering. However, this approach was limited to image-based spam detection. Moreover, Almeida et al. [16] investigated the integration of lexicographical resources, semantic dictionaries, and diverse semantic analysis techniques to refine and improve text messages, ultimately strengthening the effectiveness of the classification process.

This evolution in text classification reflects the dynamic and ever-expanding scope of NLP as it continues to explore new horizons in both theory and application.The realm of text classification, particularly in identifying spam content, has witnessed notable advances, with several researchers contributing novel techniques and methodologies that showcase the progress in machine learning and NLP. Almeida et al. [17] provided evidence that SVM classifiers had an edge over others within a new public SMS spam collection, reinforcing the relevance of SVM in text classification tasks. The traditional machine learning approaches they discussed necessitated procedures for dimensionality reduction, detailed feature engineering, and an extensive exploratory data analysis before application. Jain and Gupta [18] introduced a feature-based methodology aimed at identifying smishing attacks, focusing on distinguishing between genuine and malicious communications. Their approach underscores the evolving sophistication in detecting fraudulent content.
Ghourabi et al. [19] managed to achieve an impressive 98.37\% accuracy by harnessing the combined strengths of CNN and LSTM models, surpassing the performance of previous models. This demonstrates the effectiveness of hybrid deep learning models in handling complex pattern recognition tasks in text. Meanwhile, Bassiouni et al. [20] conducted experiments with different classifiers for the purpose of email filtering, testing against the Spambase dataset from UCI. The Random Forest (RF) algorithm stood out with a 95.45\% accuracy, while other classifiers showed competitive, albeit slightly lower, performance metrics.Abbasalizadeh et al. [57] developed PriLink, a secure link scheduling framework designed to enhance privacy in wireless networks by restricting unnecessary topology exposure while ensuring efficient resource allocation. Their approach demonstrated superior privacy preservation and faster execution compared to conventional scheduling methods. 
Saeed et al. [21] discussed the application of supervised machine learning algorithms such as J48 and KNN to segregate spam from non-spam messages, indicating that traditional classifiers still hold value in the ever-evolving landscape of spam detection. Srinivasarao et al. [22] broke new ground by developing a hybrid classifier that combines KNN and SVM. They innovated further by integrating Word2vec for feature extraction and Rat Swarm Optimization for tuning the model parameters, aiming to create a robust system for SMS spam classification. Lastly, Dharani et al. [23] proposed a model leveraging Naive Bayes coupled with TF-IDF vectorization, achieving a notable 95\% accuracy and a perfect precision score. This result exemplifies the continued relevance of Naive Bayes in text classification when paired with powerful feature extraction techniques. These varied studies and their results underscore the dynamic nature of spam detection research, where both traditional algorithms and newer, more complex models find their niches. They reflect ongoing efforts to improve accuracy, precision, and overall efficiency in text classification within the broader context of NLP.
Rayavaram et al. [55] introduced CryptoEL, a cryptography education tool for K-12 students that employs interactive simulations, AI-driven dialogues, and coding exercises to reinforce learning. Their study reported high student engagement and improved comprehension of core cryptographic concepts. The referenced studies underscore the wide-ranging applications of machine learning in categorizing spam and non-spam messages. Gautam [24] carried out a comprehensive evaluation of various machine learning classifiers for spam detection, emphasizing the distinction between spam and legitimate messages. The examined classifiers included K-Nearest Neighbors (KNN), Linear Support Vector Machine (SVM), Radial Basis Function-SVM (RBF-SVM), Random Forest, and Decision Trees, with performance metrics such as accuracy, precision, recall, and F1-score utilized for comparison. The study revealed that classifiers generally perform better on balanced datasets, as they prevent bias toward the majority class. However, KNN was found to be an exception to this trend. Notably, the Linear SVM classifier demonstrated the highest accuracy across both balanced and imbalanced datasets, highlighting its robustness and effectiveness in text classification.

Prasad et al. [36], in a 2022 study, sought to improve the accuracy of spam detection by leveraging machine learning methods, with a specific emphasis on the Multinomial Naive Bayes and Decision Tree classifiers. To ensure statistical validity, they employed GPower to determine an appropriate sample size, maintaining a power of 0.8 and an alpha threshold of 0.05. Their results revealed that the Multinomial Naive Bayes classifier attained a superior accuracy of 96.50\% compared to the Decision Tree model. Moreover, their analysis indicated no statistically significant difference in performance and loss rate between the two models, as demonstrated by a p-value exceeding 0.536. This suggests that the observed performance enhancement did not lead to additional computational burdens or overfitting

\section{Method and Materials}
\subsection{Text classification}
Text classification is a process that entails the automatic grouping of textual materials into predefined categories, utilizing the methodologies underpinned by natural language processing (NLP) [27]. It operates on a dataset of documents D = {d1, ..., dn} and is driven by a set of predetermined classes C = {c1, ..., cq}, with the aim of ascertaining the class to which a new, unclassified document belongs. The utility of text classification spans a variety of applications. It facilitates the identification of documents with shared characteristics, enables the systematic organization of documents by topic, and aids in the retrieval of documents relevant to a specific subject matter. The overarching goal of text classification is to automate the process of allocating relevant categories to documents based on their content, thereby streamlining tasks such as document management, information retrieval, and content sorting. Beyond the scope of NLP, text classification (TC) finds relevance in myriad domains. It can be instrumental in academic research for literature reviews, in legal settings for case law categorization, in digital libraries for indexing, and in the business sphere for sentiment analysis of customer feedback.
Traditional approaches to TC have underscored the significance of comprehensive dictionaries, structured sets of knowledge rules, and specialized tree kernels [28]. These methods often involve rule-based classification, where texts are categorized according to the presence or absence of specific features that have been determined by expert domain knowledge. They may also use statistical measures that consider word frequencies and distributions to assign documents to categories. While effective, traditional TC methods can be limited by their reliance on expert-crafted rules and features, which can make them labor-intensive and less adaptable to new, unseen data. With the advent of machine learning and deep learning, TC approaches have evolved significantly, with algorithms now able to learn from data, recognize complex patterns, and make predictions without explicit rule-based instructions, thereby enhancing the accuracy and scalability of text classification systems.

\subsection{NB Classification}
The Naive Bayes (NB) classifier is based on Bayes' theorem and employs a probabilistic framework, operating under the naive assumption that all features within a dataset are conditionally independent of each other. Surprisingly, the NB classifier tends to show robust performance despite this simplification, which is often violated in real-world data [29]. Particularly for smaller datasets, NB classifiers can outperform more complex models [30]. Their reliability, straightforwardness, efficiency, and accuracy contribute to their widespread usage across numerous disciplines. These models find utility in various scenarios ranging from making medical treatment decisions [31], to classifying RNA sequences in taxonomic studies [45], and to sifting through and categorizing spam in email applications [46]. Nevertheless, there can be limitations to the effectiveness of NB classifiers, especially when the assumption of independence is significantly breached and when faced with problems that require nonlinear classification approaches. The choice of a classification model should be mindful of the data characteristics and the specific challenges of the task at hand. It is advisable to compare multiple classification models, considering both computational efficiency and predictive accuracy for the dataset in question. Subsequent sections will delve into the probabilistic model of the NB classifier and its application to a straightforward example. Furthermore, we will demonstrate the training of a classifier using a publicly accessible SMS dataset, with the aim to accurately classify new, unseen messages as either spam or non-spam (ham) using Python.

\subsection{Posterior Probabilities}
To comprehend the function of Naive Bayes (NB) classifiers, one must first be acquainted with Bayes' rule. This rule is a straightforward yet potent probabilistic model conceived by Thomas Bayes. The rule is articulated as such:

\begin{equation}
P(\text { posterior })=\frac{P(\text { conditional }) \cdot P(\text { prior })}{\text { evidence }}
\end{equation}

A sphere symbolizes a randomly chosen subset, and dotted lines signify the decision limits set by classification methods. In scenarios where classes are not linearly separable, non-linear classifiers like nearest neighbor algorithms are better suited than linear ones like Naive Bayes.

\subsection{Class-Conditional Probabilities}
Bayes classifiers are used to determine whether samples follow an independent and identically distributed (i.i.d.) assumption. This indicates that each variable is independent of the others and follows the same probability distribution. A common example of an i.i.d. variable is the outcome of a coin toss, where the result of one toss has no impact on the next. Moreover, it is important to note that Naive Bayes classifiers operate under the assumption that features are conditionally independent [32]. Rather than considering all possible combinations of \( x \), class-conditional probabilities are estimated from the training data. Given a \( d \)-dimensional feature vector \( x \), the class-conditional probability is computed as follows:

\subsection{Class-Conditional Probabilities}
Bayes classifiers are employed to assess whether samples conform to the assumption of independence and identical distribution (i.i.d.). This means that each variable does not influence others and follows the same probability distribution. A common example of an i.i.d. variable is a coin toss, where the outcome of one toss does not impact subsequent tosses. Furthermore, it is important to note that Naive Bayes classifiers rely on the assumption that features are conditionally independent [32]. Instead of evaluating all possible combinations of \( x \), class-conditional probabilities are estimated using training data. Given a \( d \)-dimensional feature vector \( x \), the class-conditional probability is computed as follows:

\begin{multline}
    P\left(\mathbf{x} \mid \omega_j\right) = P\left(x_1 \mid \omega_j\right) \cdot P\left(x_2 \mid \omega_j\right) \cdot \ldots \cdot P\left(x_d \mid \omega_j\right) = \\
    \prod_{k=1}^{d} P\left(x_k \mid \omega_j\right)
\end{multline}

For categorical data, the likelihood of each attribute in the feature vector can be approximated using the maximum likelihood estimate, which essentially counts the number of occurrences:

\begin{equation}
\hat{P}\left(x_i \mid \omega_j\right) = \frac{N_{x_i, \omega_j}}{N_{\omega_j}}, \quad i = 1, \ldots, d
\end{equation}

where:
- \( N_{x_i, \omega_j} \) represents the number of instances in which feature \( x_i \) appears in samples belonging to class \( \omega_j \).
- \( N_{\omega_j} \) denotes the total count of all features within class \( \omega_j \).

Imagine a dataset of 500 documents, where 100 are identified as spam. Each document is examined based on the presence of two specific features: "hello" and "world." To estimate the probability of a new message containing "Hello World" being classified as spam, one must consider the frequency of these features within spam messages. This involves computing the conditional probability of encountering the word "world" in a known spam message and multiplying it by the conditional probability of the word "hello" appearing in a spam message:

\begin{multline}
P(\mathbf{x} = [\text{hello, world}] \mid \omega = \text{spam}) = \\P(\text{hello} \mid \text{spam}) \cdot P(\text{world} \mid \text{spam})
\end{multline}

Although the assumption of conditional independence is often violated—such as in cases where words like "peanut" and "butter" frequently co-occur in a document—Naive Bayes classifiers have nonetheless demonstrated effectiveness in a variety of applications.

\subsection{Prior Probabilities}
A prior probability is introduced as a contrast to the frequentist approach, incorporating prior knowledge or beliefs [33]:

\begin{equation}
P(\text{posterior}) = \frac{P(\text{conditional}) \cdot P(\text{prior})}{\text{evidence}}
\end{equation}

where:
- \( P(\text{spam}) \) represents the probability of any given message being spam.
- \( P(\text{normal}) \) denotes the probability of a message being non-spam [48].

When priors are equally likely, the probability of each class is computed based on observed data, factoring in both class conditions and evidence (i.e., words within a message). This differs from a frequentist or maximum likelihood approach, which considers only probabilities conditioned on the class. 

The probability of a sample belonging to a specific class can be estimated using prior knowledge from domain experts or inferred from training data. This estimation assumes that the dataset is independent, uniformly distributed, and representative of the overall population. The maximum-likelihood estimation (MLE) method is employed to derive class-conditional probabilities, applying the formula:

\begin{equation}
\hat{P}(\omega_j) = \frac{N_{\omega_j}}{N_c}
\end{equation}

where \( N_{\omega_j} \) denotes the number of samples in class \( \omega_j \), and \( N_c \) represents the total number of samples in the dataset.

In the context of spam classification, the probability of a message being classified as spam can be computed as:

\begin{equation}
\hat{P}(\text{spam}) = \frac{\text{Number of spam messages (train)}}{\text{Total training samples}}
\end{equation}

Using MLE, class-conditional probabilities are derived based on training samples, facilitating the classification process in spam detection systems[33].

\subsection{KNN Algorithm for Text Classification}

The \( k \)-Nearest Neighbors (KNN) algorithm is a widely used non-parametric method in pattern recognition and is a form of supervised learning employed for predictive classification. KNN operates by directly deriving classification rules from training data without requiring an explicit model. A test sample is classified based on the \( k \) most similar training samples, assigning it to the category that is most frequent among its \( k \)-nearest neighbors [34]. 

To classify a document \( X \) using KNN, consider a dataset with \( N \) training samples categorized into classes \( C_1, C_2, ..., C_j \). After preprocessing, each document is represented as an \( m \)-dimensional feature vector. The document \( X \) is then represented by the feature vector \( (X_1, X_2, ..., X_m) \). To classify \( X \), we compute its similarity with each training document \( d_i \), where \( d_i \) is the feature vector representation of the \( i \)-th training document:

\begin{equation}
\operatorname{SIM}(X, d_i) = \frac{\sum_{j=1}^{m} (x_j - d_{ij})}{\sqrt{\sum_{j=1}^{m} x_j^2} \cdot \sqrt{\sum_{j=1}^{m} d_{ij}^2}}
\end{equation}

To determine the probability that \( X \) belongs to a category, we select the \( k \) training samples that exhibit the highest similarity values, using:

\begin{equation}
P(X, C_j) = \sum_{d_i \in KNN} \operatorname{SIM}(X, d_i) \cdot y(d_i, C_j)
\end{equation}

where \( y(d_i, C_j) \) is a category attribute function, defined as:

\[
y(d_i, C_j) =
\begin{cases} 
1, & d_i \in C_j \\
0, & d_i \notin C_j
\end{cases}
\]

Traditional KNN classification faces three key challenges:
1. The high dimensionality of text vectors increases computational complexity.
2. The classification process depends heavily on the dataset's structure.
3. Optimizations such as dimensionality reduction, dataset balancing, and algorithm fine-tuning are necessary to enhance efficiency.

The high dimensionality of text vectors leads to significant computational complexity. To mitigate this, strategies such as dimensionality reduction, dataset downsizing, and algorithm optimization can be employed.
The method's heavy dependence on the training set for generating classification rules means that even small changes in the dataset necessitate recalculating the entire model. There is an inherent assumption that all training samples are equal in weight, which does not account for the potentially uneven distribution of the training data [35].

\subsection{Support Vector Machine (SVM) for Text Classification}

Support Vector Machines (SVMs) are robust supervised learning models rooted in the principle of structured risk minimization within computational learning theory. They are widely recognized for their efficiency in various classification tasks [52]. Due to their superior performance in text categorization benchmarks, SVMs have become the preferred classification method for managing high-dimensional datasets, which are prevalent in text-related applications. The classification process of a Support Vector Machine (SVM) is defined by a decision function:

\begin{equation}
\operatorname{ggn}: ( ( w \cdot x ) + b ).
\end{equation}

For a given dataset of instances \( (x_1, y_1), \dots, (x_l, y_l) \), where \( x \in \mathbb{R}^N \) and \( y_i \in \{-1,+1\} \) for \( i = 1, \dots, l \), the objective is to determine a decision function \( f(w,b) \) such that:

\begin{equation}
y_i ( (w \cdot x) + b ) \geq 1.
\end{equation}

In scenarios where a perfectly separable hyperplane is not feasible, slack variables \( \xi_i \) are introduced to account for misclassification, formulated as follows:

\begin{equation}
y_i ( (w \cdot x) + b ) \geq 1 - \xi_i, \quad i = 1, \dots, l.
\end{equation}

Here, \( \xi_i \) represents the slack variables, \( C \) is the penalty parameter that regulates misclassification tolerance, and \( b \) denotes the bias term. The support vectors, defined as data points \( x_i \) with non-zero Lagrange multipliers \( \alpha_i \), are critical in defining the SVM decision boundary.

\section{Linear Discriminant Analysis (LDA)}

Linear Discriminant Analysis (LDA) is a widely recognized method for both dimensionality reduction and classification. It is particularly advantageous when dealing with imbalanced class distributions. Typically, the performance of the model is assessed using randomly chosen test samples. LDA classification is grounded in variance ratio comparisons, where the total variance is contrasted with the within-class variance for dependent transformations, and the between-class variance is examined against the within-class variance for independent transformations. To determine the number of samples per category, the process follows these steps: Given $x_i\in R^d$ as $d$-dimensional input samples and $y_i \in \{1,2,\dots,c\}$ as the corresponding target labels, where $n$ signifies the document count and $c$ represents the number of classes [37].

\begin{equation}
s_W=\sum_{l=1}^c s_l
\end{equation}

The between-class scatter matrix is mathematically defined as:

\begin{equation}
S_B=\sum_{i=1}^C N_i\left(\mu_i-\mu\right)\left(\mu_i-\mu\right)^T
\end{equation}

where:

\begin{equation}
\mu=\frac{1}{N} \sum_{\forall x} x
\end{equation}

To obtain the projection of the $c-1$ vectors $w_i$ into the matrix $W$, the transformation is conducted as follows:

\begin{equation}
\begin{gathered}
W=\left[w_1\left|w_2\right| \dots \mid w_{c-1}\right] \\
y_i=w_i^T x
\end{gathered}
\end{equation}

As a result, the scatter and mean matrices are determined using the following expressions [37]:

\begin{equation}
\begin{aligned}
& \tilde{S}_W=\sum_{i=1}^c \sum_{y \in w_i}\left(y-\tilde{\mu}_i\right)\left(y-\tilde{\mu}_i\right)^T \\
& \tilde{S}_B=\sum_{i=1}^c\left(\tilde{\mu}_i-\tilde{\mu}\right)\left(\tilde{\mu}_i-\tilde{\mu}\right)^T
\end{aligned}
\end{equation}

For cases where projections exceed scalar values in $(c-1)$ dimensions, the determinant of the scatter matrix is calculated as:

\begin{equation}
J(W)=\frac{\left|S_B\right|}{\left|S_W\right|}
\end{equation}

This formulation can be restructured using the Fisher Discriminant Analysis (FDA) framework as:

\begin{equation}
J(W)=\frac{\left|W^T S_B W\right|}{\left|W^T S_W W\right|}
\end{equation}

\subsection{DT Classifier}

Decision Trees (DT) function by recursively partitioning data into subgroups through logical conditions. During the training process, the dataset is successively divided into two branches at each node based on a specific decision rule. The selection of an appropriate partitioning criterion is fundamental for constructing an optimal Decision Tree. In this study, the Gini index is employed as the primary metric for determining the structure of the decision tree, following the Classification and Regression Tree (CART) methodology [38].

\subsection{DNN Classifier}

A Deep Neural Network (DNN) is an architecture composed of multiple sequentially connected layers, where each layer receives input from the preceding layer and transmits output to the next. Connections are established across all hierarchical levels. In the output layer, each node represents a classification category, while for binary classification, a single node is sufficient. The input layer, on the other hand, consists of nodes that correspond to textual features. This structured approach forms the foundation of DNN architectures, with further details elaborated upon in subsequent sections.

\subsection{Input Layer}

The input layer, represented by $x$, consists of $n$ elements. Each component of $x$ is formulated as a $d$-dimensional dense matrix, effectively defining $x$ as a feature representation of size $d \times n$.

\subsection{Hidden Layers}

In neural networks, hidden layers play a fundamental role in feature extraction. Consider three input elements $x_1, x_2, \dots, x_n$, where the vector $c_i \in R^{wd}$ signifies the combined embedding of $w$ successive inputs $x_i - \hat{w} + 1, \dots, x_i$. Here, $w$ represents the filter width, and the condition $0 < i < w$ applies. When $i < 1$ or $i > n$, embeddings are zero-padded to ensure uniformity.

Consequently, the model representation $R^{d \times wd}$ is constructed for the $W$-gram $x_i$ when $i < 1$ or $i > n$, employing the respective convolution weights $R^{d \times wd}$ assigned to the $W$-gram. This transformation is mathematically defined as:

\begin{equation}
    p_i = \tanh (\hat{W} \cdot c_i + b)
\end{equation}

\subsection{Principal Component Analysis (PCA)}
Principal Component Analysis (PCA) is a widely used technique for reducing the dimensionality of large datasets before conducting analysis, visualization, feature selection, or data compression. Once all attributes are normalized, PCA transforms the data matrix by decomposing it into singular values or by deriving eigenvalues from the covariance matrix [40]. Typically, to improve both computational efficiency and accuracy, the data is mean-centered at the initial stage of the process.

\subsection{Feature Extraction}
Feature extraction (FE) focuses on converting textual data into a structured format, such as a keyword vector, making it more suitable for supervised learning algorithms. This technique enhances the representation of textual information by capturing important aspects, such as the frequency of key terms in a document. Identifying and encoding relevant keywords is one of the fundamental steps in supervised machine learning, as it significantly improves the classification model’s ability to recognize the most informative patterns.

\subsection{Bag-of-Words (BoW) Model for Feature Extraction}
The Bag-of-Words (BoW) method represents textual data as a set of word occurrences, disregarding grammar and the sequential order of words. For a given training pair \( (x, y) \in D \), where \( x \) is a BoW vector \( x \in \mathbb{R}^{n_{\text{vocab}}} \) and \( y \) is a categorical label \( y \in Y \), the goal is to establish a function \( \hat{y} = f_{\theta}^{(\text{BoW})}(x) \) with parameters \( \theta \), ensuring that \( \hat{y} \) approximates the actual label \( y \). When applying the BoW model with a single hidden layer, it is crucial to conduct comparative evaluations using both BoW and TF-IDF representations [41]. In multi-label classification tasks, where an instance may belong to multiple categories, binary cross-entropy is favored over categorical cross-entropy. This preference arises because class labels are treated independently, with a binary Sigmoid activation function determining the presence or absence of each label. Within the BoW framework, words are encoded as one-hot vectors. For a vocabulary of size $|\sum|$, each word is depicted as a sparse vector, where only the position corresponding to that word is set to 1, while all other elements remain 0.

 However, the scalability of the BoW model is limited due to the complexity and size of vocabularies. For example, phrases such as "This is good" and "Is this good" are assigned identical vector representations, despite differences in word arrangement, leading to a loss of syntactic information. This limitation presents challenges in computational linguistics and data science. While simple frequency-based techniques such as Term Frequency (TF) and BoW are effective for text feature extraction, they do not capture contextual relationships between words or their relative importance within a document [42].

\subsection{Evaluation of Text Classification}
When assessing the performance of document classifiers in text retrieval systems, experimental validation is generally preferred over purely analytical evaluation. This preference arises from the necessity of defining a formal problem framework to verify system accuracy and completeness. Due to the inherently subjective nature of text classification—where determining whether a document has been correctly categorized can be ambiguous—establishing a standardized evaluation criterion is challenging. Experimental assessments primarily focus on measuring classifier effectiveness in terms of prediction accuracy rather than computational efficiency.

\subsection{Performance Metrics for Classification}
The performance of a classification algorithm is assessed using several key metrics, including accuracy, recall, precision, and the F1 score. Accuracy (\( Acc_i \)) is determined by computing the ratio of correctly classified instances to the total number of instances:

\[
Acc_i = \frac{\text{Number of correctly classified texts}}{\text{Total number of texts}}
\]

Recall (\( R_i \)), also referred to as sensitivity, represents the fraction of actual relevant instances that have been successfully retrieved:

\[
R_i = \frac{\text{TP}_i}{\text{TP}_i + \text{FN}_i}
\]

Precision (\( P_i \)) quantifies the proportion of retrieved instances that are actually relevant:

\[
P_i = \frac{\text{TP}_i}{\text{TP}_i + \text{FP}_i}
\]

The F1 score serves as a harmonic mean of precision and recall, ensuring a balanced evaluation of the two metrics:

\[
F1 = 2 \cdot \frac{P_i \cdot R_i}{P_i + R_i}
\]

These performance metrics are best assessed using a confusion matrix for each class \( c_i \) within a test dataset. True positives (\( TP_i \)) represent the number of texts accurately classified as belonging to a given category, while true negatives (\( TN_i \)) correspond to texts correctly identified as not belonging to that category. False positives (\( FP_i \)) occur when a text is incorrectly assigned to a category, whereas false negatives (\( FN_i \)) occur when a text that should belong to a category is misclassified as not belonging.

\subsection{Text preprocessing}

The text preparation process typically consists of four stages: tokenization, stop word removal, stemming, and vector space modeling. Tokenization involves removing special characters and white spaces. Removing stop words is a critical step for discarding information that does not carry meaningful semantic content. According to information retrieval specialists [43], common words like "the," "a," "and," and "that," though grammatically necessary, offer little in terms of content-related significance due to their ubiquitous presence in English texts and consequently hold low discriminative value. Afterword extraction, a final check is conducted to ensure all stop words have been removed from the dataset. Stemming is a technique that simplifies words to their root forms by trimming prefixes and suffixes. This process not only reduces the keyword space but also potentially enhances text classification performance by merging variations of a word into a single representative term. Stemming is often employed in knowledge retrieval to improve efficiency and produce more pertinent search results. In vector space modeling, each term in the dataset is represented as a vector denoted by (x, td), where 'x' stands for the vector dimensions, 'R' for the terms, and 'n' for the total count of terms. The dimension of a keyword within the vector space is determined either by its frequency within the document or by its relevance to the text classification task.

\subsection{Text classification process}
The text classification process, as illustrated in Figure 1, consists of six key stages: data acquisition, text preprocessing, feature extraction, dimensionality reduction, model training, and performance assessment.

\begin{figure}[ht]
\centerline {\includegraphics[scale=0.6]{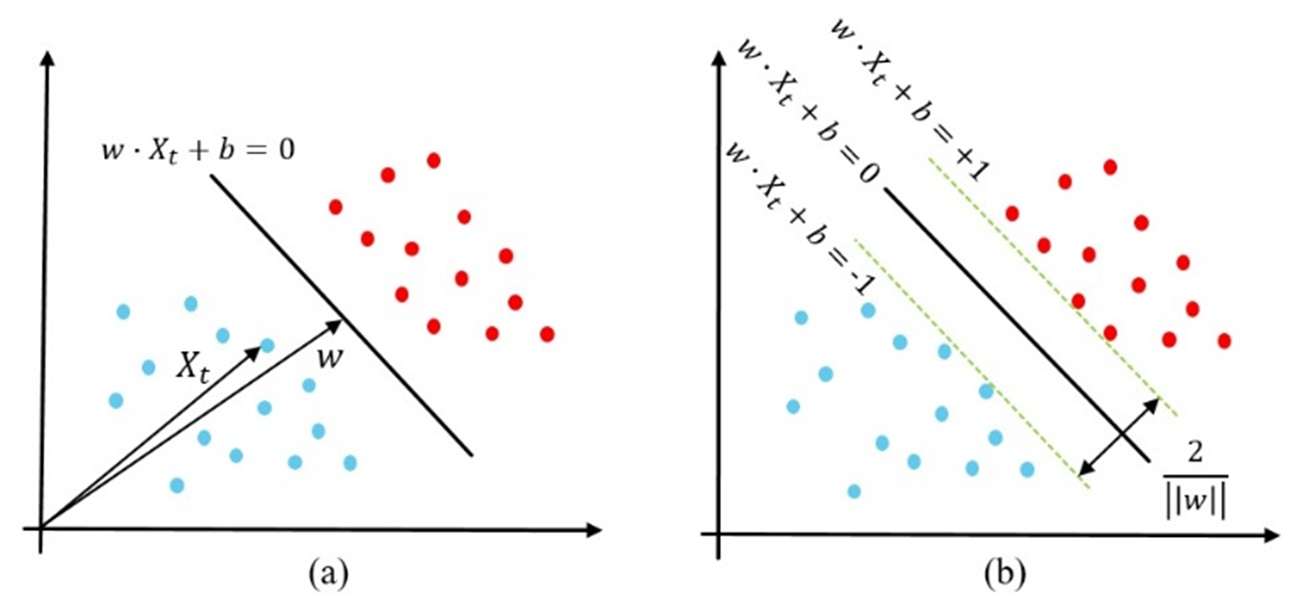}}
\caption{ Linear situations are characterized by direct proportionality between variables – as one variable changes, the other does so in a predictable, constant way. These relationships can be visually represented by a straight line on a graph.}
\label{fig1}
\end{figure}

Figure 2 presents a structured methodology for evaluating the effectiveness of various classification techniques in SMS spam detection. The process begins with data preprocessing, which entails the removal of stop words, followed by stemming and tokenization. The dataset is subsequently divided into training and test sets. Feature extraction is applied to the training data using two distinct approaches: Bag-of-Words (BoW) and Term Frequency-Inverse Document Frequency (TF-IDF). For features obtained via TF-IDF, Principal Component Analysis (PCA) is employed to reduce dimensionality to 10 principal components.

Following feature extraction, six classification models are trained on both feature sets: Naive Bayes (NB), K-Nearest Neighbors (KNN), Support Vector Machines (SVM), Linear Discriminant Analysis (LDA), Decision Trees (DT), and Deep Neural Networks (DNN). The classifiers' performance on the test set is assessed using various evaluation metrics, including precision, recall, F1-score, accuracy, and Area Under the Curve (AUC). Additionally, Receiver Operating Characteristic (ROC) curves and confusion matrices are generated for each classifier-feature combination.

This study utilizes these evaluation metrics to conduct a comparative analysis, aiming to determine which classifier exhibits the highest performance based on the obtained results. Ultimately, the study concludes with a comprehensive discussion of the findings and their broader implications.

\begin{figure}
\centerline{\includegraphics[scale=0.6]{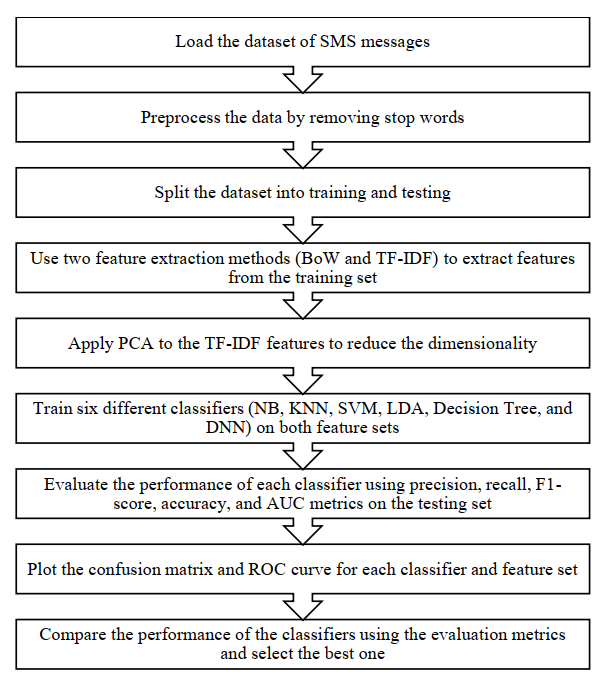}}
\caption{Conceptual diagram and workflow}
\label{fig2}
\end{figure}

\subsection{Data collection}
For our SMS spam detection project, we utilized a dataset of SMS messages available in CSV format from Kaggle. Kaggle is a well-known platform that hosts data science competitions and offers a variety of datasets [44]. The dataset comprises 5,572 records, with each entry representing an individual message. Each message in the dataset is characterized by two attributes: 'category', which specifies whether the message is spam or not, and 'message', which contains the actual text content of the SMS.. The 'message' attribute contains the text content of the SMS, while the 'category' attribute classifies each message as either 'spam' or 'ham' (the latter indicating a non-spam message).

In the preprocessing phase, the 'category' attribute was converted into a factor variable, which is essential for categorical analysis in many statistical and machine learning frameworks. The dataset exhibits an imbalance with a majority of 'ham' messages, totaling 4,825, in contrast to 750 messages labeled as 'spam'. For the purposes of training and evaluating machine learning models, the data is arranged into a matrix of dimensions 5,572 × 2, representing the observations and variables, respectively.

\section{Results and Discussion}

The primary objective of this experimental study was to design and evaluate a range of algorithms for detecting spam in SMS messages. The implementation of these algorithms was conducted using Python, PyTorch, and scikit-learn. For benchmarking purposes, we utilized well-established classification models, including Naive Bayes (NB), Support Vector Machines (SVM), k-Nearest Neighbors (KNN), Deep Neural Networks (DNN), Decision Trees (DT), and Linear Discriminant Analysis (LDA), alongside a proprietary DNN model. The performance of each classifier was assessed using multiple evaluation metrics, including precision, recall, F1-score, accuracy, and the Area Under the Receiver Operating Characteristic Curve (AUC).

Feature extraction was performed using two distinct approaches: Bag-of-Words (BoW) and Term Frequency-Inverse Document Frequency (TF-IDF). Furthermore, Principal Component Analysis (PCA) was applied to reduce the dimensionality of the TF-IDF feature space, enhancing computational efficiency. Our research also involved the deployment of a deep neural network (DNN) to classify SMS messages as either spam or non-spam. The DNN architecture consisted of six fully connected layers, with input and output dimensions corresponding to the number of features and classes, respectively. To prevent overfitting, a non-linear activation function (Tanh) was incorporated at each layer, combined with dropout regularization. The output layer of the DNN model generated the predicted class label by processing input features through the network layers. The performance of the model was evaluated on a test dataset using the same metrics as the other classifiers, namely precision, recall, F1-score, accuracy, and AUC.

A widely adopted approach for measuring classification performance is the confusion matrix, which visually represents the differences between actual and predicted class labels. In our study, confusion matrices were used to assess the effectiveness of different feature extraction methods and classification models for SMS spam detection. The confusion matrix consists of four key components:

\begin{itemize}
    \item True Positives (TP): Spam messages correctly classified as spam.
    \item False Positives (FP): Non-spam messages mistakenly classified as spam.
    \item False Negatives (FN): Spam messages incorrectly identified as non-spam.
    \item True Negatives (TN): Non-spam messages correctly identified as non-spam.
\end{itemize}

These elements were utilized to compute essential classification metrics, including accuracy, precision, recall, and the F1-score.

These components were employed to derive key classification performance metrics, including accuracy, precision, recall, and the F1-score. In this study, we assessed the effectiveness of six classification models trained using two feature extraction approaches, namely Bag-of-Words (BoW) and Term Frequency-Inverse Document Frequency (TF-IDF). The evaluation was conducted through an analysis of confusion matrices and Receiver Operating Characteristic (ROC) curves. The obtained results highlight the relative strengths and limitations of these models. The findings, along with the corresponding confusion matrices, are depicted in Figures 3 and 6.

\begin{figure*}
\centerline{\includegraphics[scale=0.4]{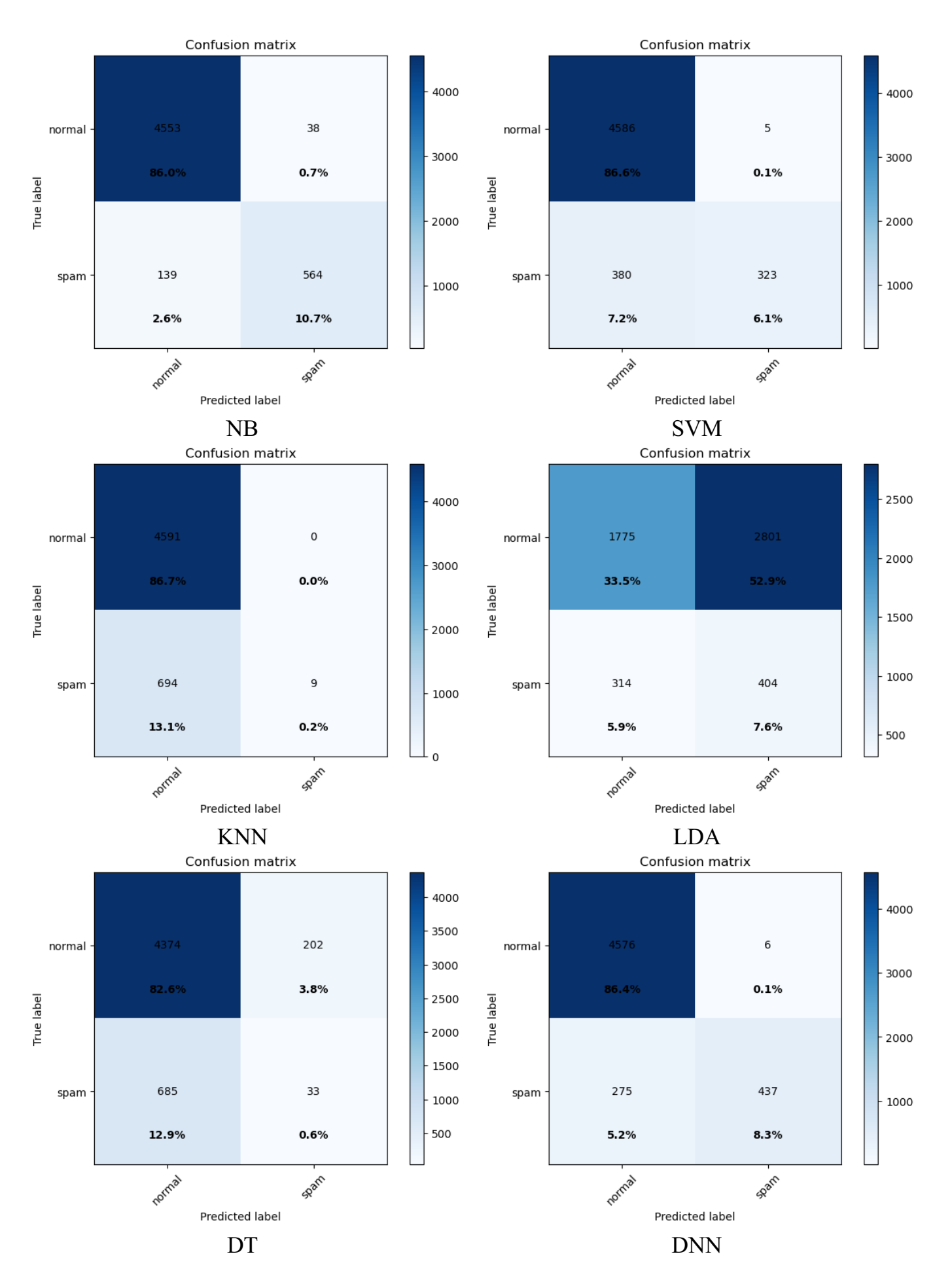}}
\caption{The confusion matrices for various machine learning models utilizing the Bag-of-Words (BoW) feature extraction technique}
\label{fig3}
\end{figure*}

The Receiver Operating Characteristic (ROC) curve is a fundamental tool for evaluating the performance of classification models. It visually illustrates the trade-off between the True Positive Rate (TPR) and the False Positive Rate (FPR) across varying threshold levels. The TPR indicates the proportion of correctly classified positive instances, while the FPR represents the fraction of negative instances incorrectly classified as positives. The ROC curve is generated by systematically adjusting the classification threshold and plotting the resulting TPR and FPR values. Figure 4, in this study, provides a comparative evaluation of different classifiers employing the Bag-of-Words (BoW) method for feature extraction.

\begin{figure}
\centerline{\includegraphics[scale=0.4]{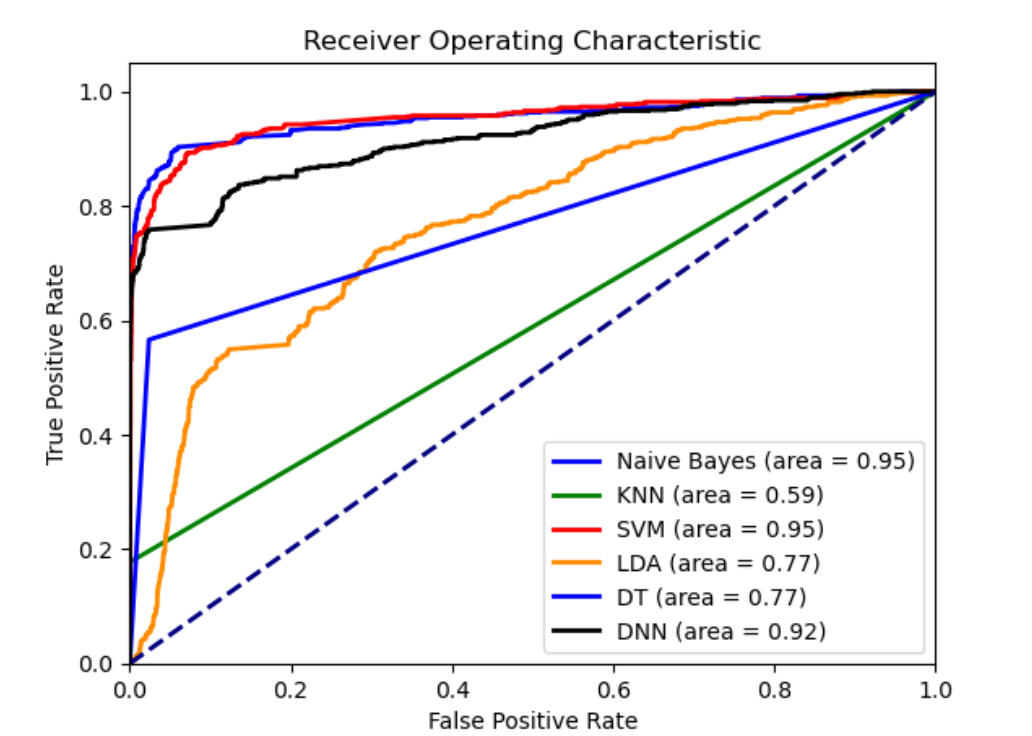}}
\caption{The ROC curve comparing different machine learning models employing the Bag-of-Words (BoW) feature extraction method}
\label{fig4}
\end{figure}

Referring to Figure 4, the Area Under the Curve (AUC) values for the six classifiers employing the Bag-of-Words (BoW) feature extraction method are analyzed. As illustrated in the figure, the Naive Bayes (NB) classifier attains the highest AUC score with BoW, achieving values of 0.96 for non-spam messages and 0.95 for spam messages. In contrast, the k-Nearest Neighbors (KNN) classifier records the lowest AUC values for the BoW approach, with 0.87 for non-spam and 0.59 for spam messages. The overall range of AUC values for non-spam messages using BoW spans from 0.87 to 0.96, whereas for spam messages, the values vary between 0.59 and 0.95. Among the evaluated classifiers, the NB model consistently outperforms others when utilizing this particular feature extraction technique.

\begin{figure}
\centerline{\includegraphics[scale=0.44]{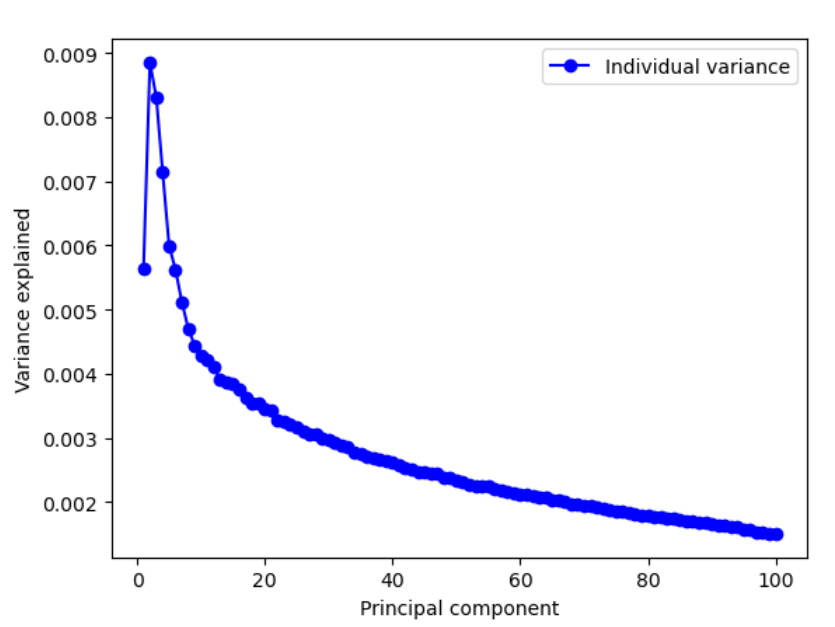}}
\caption{The scree plot illustrating the feature reduction of TF-IDF features using Principal Component Analysis (PCA)}
\label{fig5}
\end{figure}

The Scree plot in Figure 4 depicts the proportion of variance in the dataset explained by each principal component obtained from the Principal Component Analysis (PCA) applied to the TF-IDF features. The plot demonstrates a declining trend, indicating that as the number of principal components increases, the amount of variance they capture diminishes. This suggests that the first few principal components retain the most crucial information necessary for the classification process. In this study, the top 10 principal components, which collectively accounted for a significant portion of the total variance, were chosen as the feature set for the classification models.

\begin{figure*}
\centerline{\includegraphics[scale=0.4]{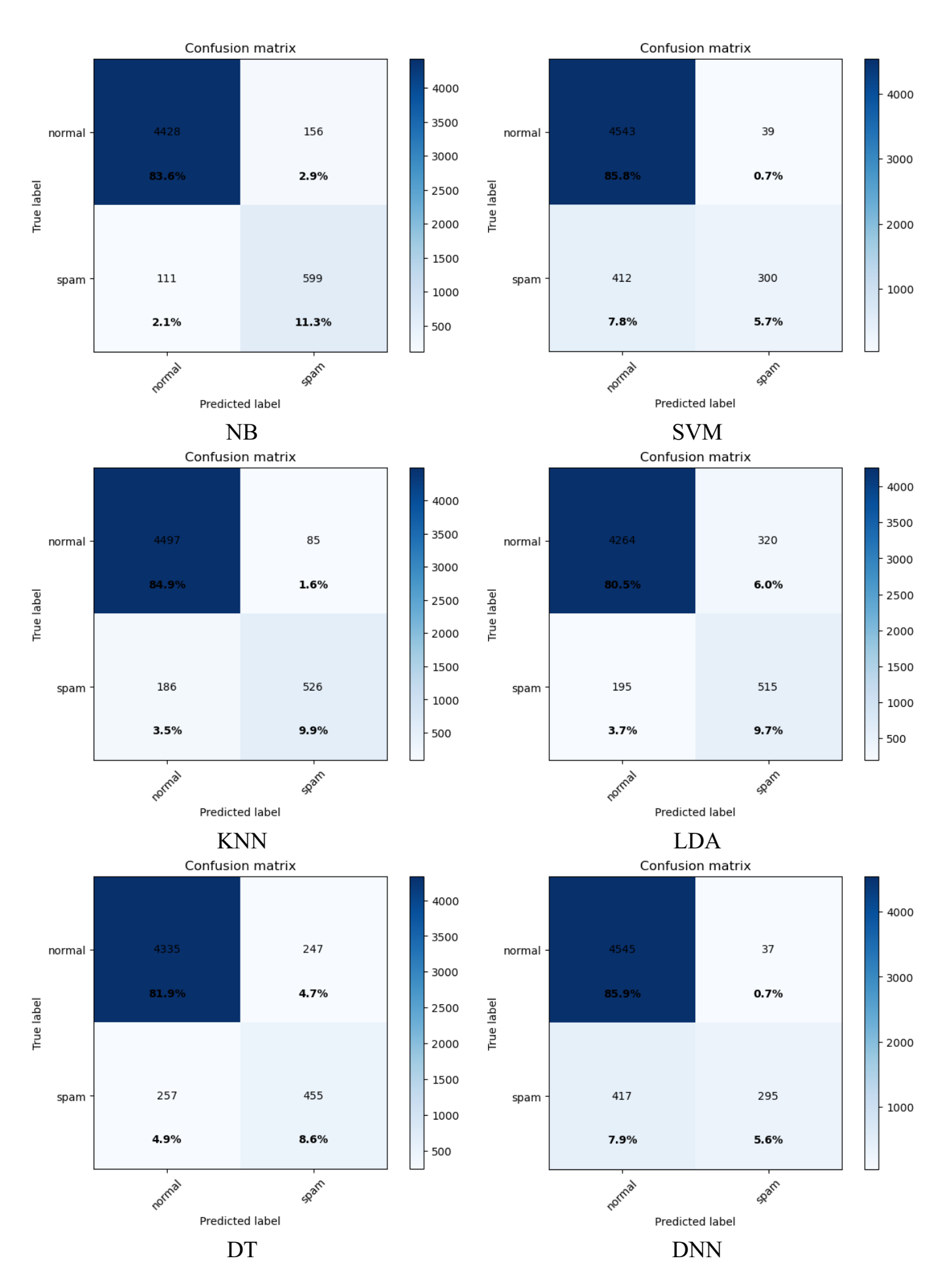}}
\caption{The confusion matrices for various machine learning models utilizing the Term Frequency-Inverse Document Frequency (TF-IDF) feature extraction technique}
\label{fig6}
\end{figure*}

For the Naive Bayes classifier, the Term Frequency-Inverse Document Frequency (TF-IDF) feature extraction method exhibited the highest performance, achieving an accuracy of 0.97 and a recall of 0.75 for spam messages. The model maintained a high overall accuracy of 0.97, with an F1-score of 0.84. Furthermore, the Area Under the Curve (AUC) value of 0.96 highlighted the classifier's robust ability to distinguish between spam and non-spam messages. 

In contrast, the Bag-of-Words (BoW) feature extraction method yielded the best results with the k-Nearest Neighbors (KNN) classifier, attaining an accuracy of 0.87 and a perfect recall of 1 for non-spam messages. However, its capability in detecting spam messages was significantly lower, with an accuracy of only 0.04. The corresponding F1-score of 0.59 and an AUC of 0.59 indicated suboptimal overall performance. For the Support Vector Machine (SVM) classifier, the TF-IDF feature extraction method demonstrated strong effectiveness, with an accuracy of 0.92 and a recall of 0.88 for spam messages. The model's overall accuracy was recorded at 0.95, while the F1-score reached 0.42. Additionally, an AUC of 0.96 further validated the classifier's reliability in spam detection.

\begin{figure}
\centerline{\includegraphics[scale=0.45]{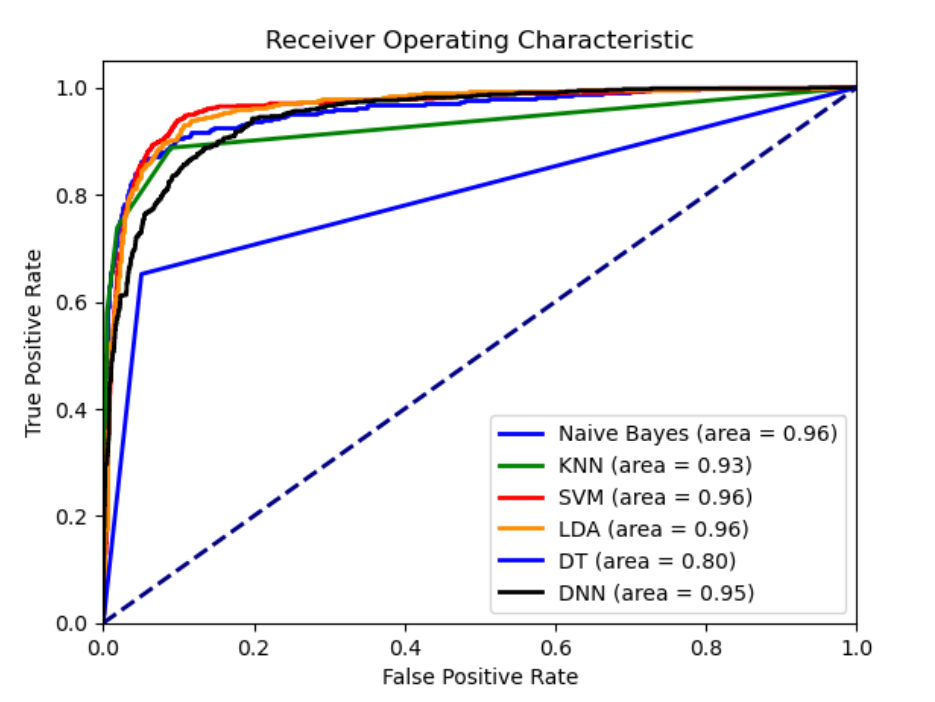}}
\caption{The ROC curve comparing the performance of different machine learning models using the TF-IDF feature extraction technique}
\label{fig7}
\end{figure}

Regarding Figure 6, the Area Under the Curve (AUC) values for the six classifiers employing the Term Frequency-Inverse Document Frequency (TF-IDF) feature extraction technique exhibited variations across different models. The Deep Neural Network (DNN) classifier achieved the highest AUC score of 0.95, demonstrating its capability in effectively distinguishing between spam and non-spam messages. Similarly, the Naive Bayes (NB) and Linear Discriminant Analysis (LDA) classifiers reported strong AUC values of 0.94 and 0.96, respectively. Furthermore, the Support Vector Machine (SVM) classifier attained the second-highest AUC value, as detailed in Table 2.

\begin{table*}[h]
\centering
\caption{Comparison of Machine Learning Methods Based on Performance Metrics}
\resizebox{\textwidth}{!}{%
\begin{tabular}{|c|c|c|c|c|c|c|c|c|c|}
\hline
\textbf{Classification Model} & \textbf{Feature Extraction} & \textbf{Precision (N)} & \textbf{Precision (S)} & \textbf{Recall (N)} & \textbf{Recall (S)} & \textbf{F1-score (N)} & \textbf{F1-score (S)} & \textbf{Overall Accuracy} & \textbf{AUC} \\
\hline
Naive Bayes  & Bag-of-Words   & 0.971 & 0.939 & 0.994 & 0.781 & 0.985 & 0.861 & 0.965 & 0.953 \\
             & TF-IDF & 0.976 & 0.754 & 0.968 & 0.849 & 0.970 & 0.796 & 0.946 & 0.962 \\
\hline
K-Nearest Neighbors & Bag-of-Words   & 0.874 & 1.002 & 1.004 & 0.048 & 0.931 & 0.101 & 0.872 & 0.594 \\
                    & TF-IDF & 0.962 & 0.865 & 0.985 & 0.741 & 0.973 & 0.808 & 0.953 & 0.931 \\
\hline
Support Vector Machines & Bag-of-Words   & 0.923 & 0.995 & 1.000 & 0.459 & 0.961 & 0.627 & 0.926 & 0.952 \\
                        & TF-IDF & 0.926 & 0.891 & 0.990 & 0.429 & 0.953 & 0.571 & 0.914 & 0.961 \\
\hline
Linear Discriminant Analysis & Bag-of-Words   & 0.924 & 0.423 & 0.884 & 0.538 & 0.909 & 0.474 & 0.845 & 0.769 \\
                             & TF-IDF & 0.965 & 0.811 & 0.972 & 0.728 & 0.976 & 0.765 & 0.945 & 0.963 \\
\hline
Decision Trees  & Bag-of-Words   & 0.948 & 0.798 & 0.987 & 0.578 & 0.968 & 0.663 & 0.924 & 0.770 \\
                & TF-IDF & 0.943 & 0.656 & 0.953 & 0.642 & 0.653 & 0.643 & 0.899 & 0.803 \\
\hline
Deep Neural Network & Bag-of-Words   & 0.943 & 0.997 & 1.003 & 0.612 & 0.973 & 0.764 & 0.954 & 0.923 \\
                    & TF-IDF & 0.928 & 0.890 & 0.991 & 0.415 & 0.952 & 0.567 & 0.910 & 0.954 \\
\hline
\end{tabular}%
}
\label{table:ml_comparison}
\end{table*}

According to Table 2, the Term Frequency-Inverse Document Frequency (TF-IDF) feature extraction method achieved the best classification performance when combined with Linear Discriminant Analysis (LDA). Specifically, for spam detection, the model recorded a precision of 0.81 and a recall of 0.97, reflecting strong classification capability. The model attained an overall accuracy of 0.97, with an F1-score of 0.72. Additionally, an Area Under the Curve (AUC) value of 0.94 confirmed its effectiveness in distinguishing between spam and non-spam messages. On the other hand, using the Bag-of-Words (BoW) feature extraction method alongside Decision Trees (DT) led to relatively lower classification performance. For spam classification, this approach yielded a precision of 0.65 and a recall of 0.95. However, the overall accuracy was only 0.65, with an F1-score of 0.64. The corresponding AUC score for this combination was 0.80, indicating moderate performance. Despite these results, the BoW method produced the best outcomes when applied to the Deep Neural Network (DNN) classifier, achieving a precision of 0.89 and a recall of 0.99 for spam messages. The model maintained an accuracy of 0.95, although the F1-score was relatively lower at 0.41. With an AUC of 0.92, the overall classification performance of this model was deemed satisfactory.

Overall, the findings suggest that the TF-IDF feature extraction technique consistently surpasses the BoW method across all six classification models assessed. In particular, TF-IDF demonstrated significant effectiveness when paired with Naive Bayes (NB), Support Vector Machines (SVM), and Deep Neural Networks (DNN). In contrast, classifiers such as k-Nearest Neighbors (KNN), Linear Discriminant Analysis (LDA), and Decision Trees (DT) exhibited lower effectiveness, regardless of the feature extraction technique used.

\section{Conclusion }

The findings of this study reveal that among the six classification models examined, the Term Frequency-Inverse Document Frequency (TF-IDF) feature extraction method consistently outperforms the Bag-of-Words (BoW) approach for detecting spam in SMS messages. The evaluation was carried out using various performance metrics, including precision, recall, F1-score, accuracy, and the Area Under the Curve (AUC), to assess the effectiveness of each classification model.

The TF-IDF technique demonstrated the highest performance when paired with the Naive Bayes (NB) classifier, achieving a precision of 0.97 and a recall of 0.75 for spam messages. Likewise, the Support Vector Machine (SVM) classifier exhibited strong performance with TF-IDF, recording an accuracy of 0.92 and a recall of 0.88 for spam classification. In contrast, classifiers such as Linear Discriminant Analysis (LDA) and Decision Trees (DT) showed comparatively lower effectiveness, regardless of the feature extraction method applied. When employing the BoW technique, the k-Nearest Neighbors (KNN) classifier performed well in recognizing non-spam messages but faced challenges in accurately identifying spam content.  These findings suggest that integrating the TF-IDF feature extraction approach with classifiers such as NB, SVM, or Deep Neural Networks (DNN) presents a reliable strategy for spam detection in SMS messages.

 The Deep Neural Network (DNN) model showed solid performance overall, with BoW providing the most favorable results for this classifier. In conclusion, the study suggests that integrating the TF-IDF feature extraction method with NB, SVM, or DNN classifiers is an effective strategy for identifying spam in user SMS messages. Caution is advised when considering KNN, LDA, and DT for this task. The use of confusion matrices and ROC curves can guide future researchers in selecting the most effective feature extraction and classification techniques for SMS spam detection. Further research might explore the performance of other feature extraction methods, like word embedding or character n-grams. It could also investigate the potential benefits of employing ensemble methods and active learning strategies to enhance SMS spam detection capabilities.

\section{Limitations}

Despite the promising results obtained in this study, there are several limitations that should be acknowledged:

\begin{itemize}
    \item The dataset used in this study contained a disproportionate number of non-spam messages compared to spam messages. This imbalance could affect the generalizability of the results, potentially leading to biased classifier performance.
    \item Feature Representation Constraints: While TF-IDF outperformed BoW, both methods primarily rely on word frequency and do not fully capture the semantic meaning or contextual relationships between words.
    \item  Some models, such as KNN and DT, exhibited lower performance for spam detection. These models may not be well-suited for handling large-scale text classification problems and require further optimization.
    \item  The study was conducted using a single publicly available dataset. The results may vary when applied to different datasets with varying linguistic structures, requiring further validation.
    \item The study did not explore modern deep learning-based language models, such as word embeddings (Word2Vec, GloVe) or transformer-based models (BERT), which could potentially enhance classification accuracy.
\end{itemize}

\section{Future Work}

While this study has demonstrated the effectiveness of the TF-IDF feature extraction method in conjunction with various classifiers, there remain several avenues for future research. One potential direction is the integration of contextual word embeddings, such as BERT, Word2Vec, or fastText, to better capture semantic relationships and improve classification accuracy. Additionally, addressing the dataset imbalance issue through techniques like data augmentation, synthetic data generation, or resampling methods (e.g., SMOTE) could enhance the robustness of spam detection models. Future studies may also explore hybrid approaches by combining multiple feature extraction techniques, such as TF-IDF with deep learning-based embeddings, to improve classification performance. Moreover, expanding the dataset to include multilingual SMS messages could provide insights into the effectiveness of classifiers across different languages and linguistic structures. Another important consideration is the implementation of real-time spam filtering systems and evaluating their computational efficiency in production environments.

\section{References}
1.	Annamoradnejad I, Habibi J, Fazli M. Multi-view approach to suggest moderation actions in community question answering sites. Information Sciences. 2022 Jul 1;600:144-54.\\
2.	Abkenar SB, Kashani MH, Akbari M, Mahdipour E. Twitter spam detection: a systematic review. arXiv preprint arXiv:2011.14754. 2020 Nov 30.\\
3.	Saeidnia HR, Hosseini E, Lund B, Tehrani MA, Zaker S, Molaei S. Artificial intelligence in the battle against disinformation and misinformation: a systematic review of challenges and approaches. Knowledge and Information Systems. 2025 Jan 27:1-20.\\
4.	Sharma R, Singh B, Khamparia A. Machine Learning and Generative AI Techniques for Sentiment Analysis with Applications. Generative Artificial Intelligence for Biomedical and Smart Health Informatics. 2025 Jan 9:183-208.\\
5.	Kowsari K, Jafari Meimandi K, Heidarysafa M, Mendu S, Barnes L, Brown D. Text classification algorithms: A survey. Information. 2019 Apr 23;10(4):150.\\
6.	Kahl S, Wood CM, Eibl M, Klinck H. BirdNET: A deep learning solution for avian diversity monitoring. Ecological Informatics. 2021 Mar 1;61:101236.\\
7.	Thielmann A, Weisser C, Krenz A, Säfken B. Unsupervised document classification integrating web scraping, one-class SVM and LDA topic modelling. Journal of Applied Statistics. 2023 Feb 17;50(3):574-91.\\
8.	Alodadi N. Characterization, Extraction, and Impact Assessment of Semantic Aspects in Online Physician Reviews (Doctoral dissertation, University of Maryland, Baltimore County).\\
9.	Jang J, Kim Y, Choi K, Suh S. Sequential targeting: a continual learning approach for data imbalance in text classification. Expert Systems with Applications. 2021 Oct 1;179:115067.\\
10.	P.K. Roy, J.P. Singh and S. Banerjee, "Deep learning to filter SMS Spam", Future Gener. Comput. Syst, vol. 102, pp. 524-533, 2020.\\
11.	J.W. Joo, S.Y. Moon, S. Singh and J.H. Park, "S-Detector: an enhanced security model for detecting Smishing attack for mobile computing", Telecommun. Syst, vol. 66, pp. 29-38, 2017.\\
12.	Chandra A, Khatri SK (2019a) Spam SMS filtering using recurrent neural network and long short term memory. In 2019 4th international conference on information systems and computer networks (ISCON) (pp. 118-122).\\
13.	Chandra A, Khatri SK (2019b) Spam SMS filtering using recurrent neural network and long short term memory. In 2019 4th international conference on information systems and computer networks (ISCON) (pp. 118-122). \\
14.	Lee HY, Kang SS (2019) Word embedding method of sms messages for spam message filtering. In 2019 IEEE.\\ international conference on big data and smart computing (BigComp) (pp. 1-4). IEEE.\\
15.	Q. Xu, E. W. Xiang, Q. Yang, J. Du, and J. Zhong, "SMS Spam Detection Using Noncontent Features," IEEE Intell. Syst., vol. 27, no. 6, pp. 44– 51, Nov. 2012, doi: 10.1109/MIS.2012.3 
16.	T. A. Almeida, J. M. G. Hidalgo, and A. Yamakami. "Contributions to the study of SMS spam filtering: new collection and results," in Proc. of the 11th ACM symposium on Document engineering, pp. 259-262, 2011, doi: 10.1145/2034691.2034742.\\
17.	T. A. Almeida, J. M. G. Hidalgo, and A. Yamakami. "Contributions to the study of SMS spam filtering: new collection and results," in Proc. of the 11th ACM symposium on Document engineering, pp. 259-262, 2011, doi: 10.1145/2034691.2034742.\\
18.	Jain,  A.  K.,  \&  Gupta,  B.  B.  (2019).  Feature-based  approach  for  detection  of  smishing  messages  in  the  mobile environment.Journal of Information Technology Research (JITR),12(2), 17-3.\\
19.	Ghourabi, A., Mahmood, M. A., \& Alzubi, Q. M. (2020). A hybrid CNN-LSTM model for SMS spam detection in Arabic and english messages.Future Internet,12(9), 156.\\
20.	Bassiouni,  M.,  Ali,  M.,  \&  El-Dahshan,  E.  A.  (2018).  Ham  and  spam  e-mails  classification  using  machine  learning techniques.Journal of Applied Security Research,13(3), 315-33.\\
21.	Saeed VA. A Method for SMS Spam Message Detection Using Machine Learning.\\
22.	Srinivasarao U, Sharaff A. Machine intelligence based hybrid classifier for spam detection and sentiment analysis of SMS messages. Multimedia Tools and Applications. 2023 Feb 23:1-31.\\
23.	Dharani V, Hegde D. Spam SMS (or) Email Detection and Classification using Machine Learning. In2023 5th International Conference on Smart Systems and Inventive Technology (ICSSIT) 2023 Jan 23 (pp. 1104-1108). IEEE.\\
24.	Gautam S. Comparison of Feature Representation Schemes to Classify SMS Text using Data Balancing.\\
25.	Raga S, BL MC. Machine Learning and Deep Learning Techniques for SMS Spam Detection, Accuracy Check and Comparative Study. Journal homepage: www. ijrpr. com ISSN.;2582:7421.\\
26.	Prasad JK, Christy S. SMS Spam Detection Using Multinational Naive Bayes Algorithm Compared with Decision Tree Algorithm. Baltic Journal of Law \& Politics. 2022 Dec 13;15(4):349-56.\\
27.	Kadhim AI. Survey on supervised machine learning techniques for automatic text classification. Artificial Intelligence Review. 2019 Jun 1;52(1):273-92.\\
28.	Sebastiani F. Machine learning in automated text categorization. ACM computing surveys (CSUR). 2002 Mar 1;34(1):1-47.\\
29.	Al-Aidaroos KM, Bakar AA, Othman Z. Naive Bayes variants in classification learning. In2010 international conference on information retrieval \& knowledge management (CAMP) 2010 Mar 17 (pp. 276-281). IEEE.\\
30.	Hasan N, Choudhary S, Naaz N, Sharma N, Laskar RA. Recent advancements in molecular marker-assisted selection and applications in plant breeding programmes. Journal of Genetic Engineering and Biotechnology. 2021 Dec;19(1):1-26.\\
31.	Sanz EP, Hidalgo JM, Pérez JC. Email spam filtering. Advances in computers. 2008 Jan 1;74:45-114.\\
32.	Nechyba MC, Xu Y. Stochastic similarity for validating human control strategy models. IEEE Transactions on Robotics and Automation. 1998 Jun;14(3):437-51.\\
33.	Raschka S. Naive bayes and text classification i-introduction and theory. arXiv preprint arXiv:1410.5329. 2014 Oct 16.\\
34.	Sethi R, Kaushik I. Hand written digit recognition using machine learning. In2020 IEEE 9th International Conference on Communication Systems and Net\\work Technologies (CSNT) 2020 Apr 10 (pp. 49-54). IEEE.\\
35.	Diao R, Shen Q. Feature selection with harmony search. IEEE Transactions on Systems, Man, and Cybernetics, Part B (Cybernetics). 2012 May 23;42(6):1509-23.\\
36.	Christmann A, Steinwart I. On robustness properties of convex risk minimization methods for pattern recognition. The Journal of Machine Learning Research. 2004 Dec 1;5:1007-34.\\
37.	Sugiyama, M. Dimensionality reduction of multimodal labeled data by local fisher discriminant analysis. J. Mach. Learn. Res. 2007, 8, 1027–1061.\\
38.	Sebastiani F. Machine learning in automated text categorization. ACM computing surveys (CSUR). 2002 Mar 1;34(1):1-47.\\
39.	Liang B, Li H, Su M, Bian P, Li X, Shi W. Deep text classification can be fooled. arXiv preprint arXiv:1704.08006. 2017 Apr 26.\\
40.	Bro R, Smilde AK. Principal component analysis. Analytical methods. 2014;6(9):2812-31.\\
41.	Uysal AK, Gunal S, Ergin S, Gunal ES. The impact of feature extraction and selection on SMS spam filtering. Elektronika ir Elektrotechnika. 2013 May 1;19(5):67-72.\\
42.	Zheng H, Cheng G, Li Y, Liu C. A new fault diagnosis method for planetary gear based on image feature extraction and bag-of-words model. Measurement. 2019 Oct 1;145:1-3.\\
43.	Tijani OD, Akinwale AT, Onashoga A, Adeleke EO. AN IMPROVED AGGREGATED METHOD FOR GENERATION OF YORUBA LANGUAGE STOP WORDS (Doctoral dissertation, Department of Computer Science, College of Physical Sciences, Federal University of Agriculture, Abeokuta).\\
44.	Spam Text Message Classification. (n.d.). Spam Text Message Classification | Kaggle. https:///datasets/team-ai/spam-text-message-classification\\
45. Shen, Linjie, Yanbin Wang, Zhao Li, and Wenrui Ma. "SMS Spam Detection Using BERT and Multi-Graph Convolutional Networks." Available at SSRN 5078344.\\
46.Haider Rizvi, Syed Mustafa, Ramsha Imran, and Arif Mahmood. "Text Classification Using Graph Convolutional Networks: A Comprehensive Survey." ACM Computing Surveys (2025).\\
47.Tusher, E.H., Ismail, M.A. and Mat Raffei, A.F., 2025. Email Spam Classification Based on Deep Learning Methods: A Review. Iraqi Journal for Computer Science and Mathematics, 6(1), p.2.\\
48.Gong, Jingzhi, Vardan Voskanyan, Paul Brookes, Fan Wu, Wei Jie, Jie Xu, Rafail Giavrimis, Mike Basios, Leslie Kanthan, and Zheng Wang. "Language models for code optimization: Survey, challenges and future directions." arXiv preprint arXiv:2501.01277 (2025).\\
49.Liu, Z., 2024, April. A review of advancements and applications of pre-trained language models in cybersecurity. In 2024 12th International Symposium on Digital Forensics and Security (ISDFS) (pp. 1-10). IEEE.\\
50.Chen, Yiren, Mengjiao Cui, Ding Wang, Yiyang Cao, Peian Yang, Bo Jiang, Zhigang Lu, and Baoxu Liu. "A survey of large language models for cyber threat detection." Computers \& Security (2024): 104016.\\
51.Alshattnawi, Sawsan, Amani Shatnawi, Anas MR AlSobeh, and Aws A. Magableh. "Beyond Word-Based Model Embeddings: Contextualized Representations for Enhanced Social Media Spam Detection." Applied Sciences 14, no. 6 (2024): 2254.\\
52.Chataut, Robin, Aadesh Upadhyay, Yusuf Usman, Mary Nankya, and Prashnna K. Gyawali. "Spam No More: A Cross-Model Analysis of Machine Learning Techniques and Large Language Model Efficacies." In 2024 8th Cyber Security in Networking Conference (CSNet), pp. 116-122. IEEE, 2024.\\
53.Samad, Saleem Raja Abdul, Pradeepa Ganesan, Justin Rajasekaran, Madhubala Radhakrishnan, Hariraman Ammaippan, and Vinodhini Ramamurthy. "SmishGuard: Leveraging Machine Learning and Natural Language Processing for Smishing Detection." International Journal of Advanced Computer Science \& Applications 14, no. 11 (2023).\\
54.Shrestha, Neeraj. "A Novel Spam Email Detection Mechanism Based on XLNet." Master's thesis, University of Toledo, 2023.\\
55.Rayavaram, Pranathi, Ukaegbu Onyinyechukwu, Maryam Abbasalizadeh, Krishnaa Vellamchetty, and Sashank Narain. "CryptoEL: A Novel Experiential Learning Tool for Enhancing K-12 Cryptography Education." arXiv preprint arXiv:2411.02143 (2024).\\
56.Yao, Jinliang, Chenrui Wang, Chuang Hu, and Xiaoxi Huang. "Chinese spam detection using a hybrid BiGRU-CNN network with joint textual and phonetic embedding." Electronics 11, no. 15 (2022): 2418.\\
57- Abbasalizadeh, Maryam, Jeffrey Chan, Pranathi Rayavaram, Yimin Chen, and Sashank Narain. "Privacy-preserving link scheduling for wireless networks." IEEE Access (2024).\\
58-amad, Saleem Raja Abdul, Sundaravadivazhagan Balasubramaniyan, Pradeepa Ganesan, Amna Salim Al-Kaabi, Hariraman Ammaippan, and Jeyakumar Manickam Sam. "SMSecure: Leveraging Machine Learning for Smishing Detection." In International Conference on Data Engineering and Machine Intelligence, pp. 257-271. Singapore: Springer Nature Singapore, 2023.\\

\end{document}